# Classification Accuracy and Parameter Estimation in Multilevel Contexts: A Study of Conditional Nonparametric Multilevel Latent Class Analysis

Chi Chang[1, 2], Kimberly S. Kelly[3], M. Lee Van Horn[4], Richard T. Houang[5], Joseph Gardiner[2], Laurie Van Egeren[6], Heng-Chieh Wu[6]

[1] Office of Medical Education Research and Development, College of Human Medicine, Michigan State University, USA

[2] Department of Epidemiology and Biostatistics, College of Human Medicine, Michigan State University, USA

[3] Measurement and Quantitative Methods Program, College of Education, Michigan State University, USA.

[4] Methodology Group, College of Education, University of New Mexico, USA

[5] Center for the Study of Curriculum, College of Education, Michigan State University, USA

[6] Office of University Outreach and Engagement, Michigan State University, USA

**Abstract**

The current research has two aims. First, to demonstrate the utility conditional nonparametric multilevel latent class analysis (NP-MLCA) for multi-site program evaluation using an empirical dataset. Second, to investigate how classification accuracy and parameter estimation of a conditional NP-MLCA are affected by six study factors: the quality of latent class indicators, the number of latent class indicators, level-1 covariate effects, cross-level covariate effects, the number of level-2 units, and the size of level-2 units. A total of 96 conditions was examined using a simulation study. The resulting classification accuracy rates, the power and type-I error of cross-level covariate effects and contextual effects suggest that the nonparametric multilevel latent class model can be applied broadly in multilevel contexts.

*Keyword:* multilevel latent class analysis, mixture modeling, simulation studies, cross-level effect, conditional multilevel mixture model

# 1. Introduction

Program evaluators have long been interested in finding rigorous methods to identify better performing programs and successful program participants, and in helping underperforming participants to improve. The advantage of nonparametric multilevel latent class analysis (NP-MLCA) is that it can simultaneously identify typologies of students within schools and of the schools themselves (Chang, 2016; Finch & French, 2013; Lukociene, Varriale, & Vermunt, 2010; Vermunt, 2003). It has been applied in education to identify academic achievement and socio-economic status typologies in schools around the world using Programme for International Student Assessment data (Finch & Marchant, 2013), to identify healthy schools based on students' healthy behavior and well-being (Allison, Adlaf, Irving, Schoueri-Mychasiw, & Rehm, 2016), and to identify typologies for teacher preparation programs and future mathematics teachers based on Opportunity to Learn indicators using the Teacher Education and Development Study in Mathematics dataset (Chang, 2017). The model has been applied in epidemiology to evaluate diagnostics for trachoma (Koukounari et al., 2013); in public health to identify types of food retail environments from consumers' dietary knowledge (Zhang, Van Der Lans, & Dagevos, 2012); and in organizational research to identify categories of departments based on the proportions of four types of job demands and job control (Mäkikangas et al., 2018).

Previous methodological studies in NP-MLCA include, but are not limited to, investigating analytic relationship between nonparametric and parametric multilevel latent class analysis in terms of discrete approximation of intraclass correlation, random coefficient distributions, and residual heteroscedasticity (Rights & Sterba, 2016), power and type I error of local fit statistics (Nagelkerke, Oberski, & Vermunt, 2017), and classification enumeration (Lukočien & Vermunt, 2010) of NP-MLCA. While covariates at different levels are taken into account in the model (hereafter referred to as conditional NP-MLCA), there are few studies regarding how the effects of covariates and the quality of indicators affect classification accuracy at both the individual and school levels. The current research aims to advance our knowledge of NP-MLCA's classification performance and covariate estimation performance in cases where numerous indicators and covariates at different levels are available.

When classification is the primary interest of the researcher, cluster analysis has traditionally been applied. It uses distance measures, such as nearest-neighbor classifiers, to identify clusters among the measured or observed variables (Hair, Black, Babin, Anderson, & Tatham, 2010). However, several limitations of this method restrict its application (Chang, 2016). First, because the clusters it identifies may differ greatly depending on the distance measures selected, the method is applied mainly for exploratory purposes. Second, the way the observed indicators are ordered can have an impact on clustering results. Third, if a case is dropped, the clustering may change drastically– a further indication of the oversensitivity of the method. Fourth, cutting the hierarchical clustering tree (i.e., dendrogram) at a given height affects the classification precision and the solution is not easily rectified. In short, the results of clustering are often sensitive and unstable due to their reliance on distance measures, and for this reason, a more satisfactory solution is necessary.

Moreover, classification error cannot be incorporated into subsequent analyses of the clusters that the method produces. If researchers want to investigate further the characteristics of the clusters by examining the relationship between covariates and the cluster variable, or by examining the effect of clusters on an outcome variable (such as achievement), the usual strategy is to treat the estimated cluster membership as the observed variable and conduct a relational analysis as the second step. But again, because different distance measures can yield very different clustering solutions, the subsequent analyses are likely to result in widely divergent statistical inferences.

NP-MLCA is a method for identifying latent classes (i.e., human subgroups or types that are not identifiable except via their pattern of responses to a given set of indicators) at multiple levels simultaneously. The higher-level classification in NP-MLCA is based on which patterns of latent classes (e.g., smokers, bullying victims) are prevalent in each higher-level unit (e.g., school) (Henry & Muthén, 2010a; Mawditt, Sacker, Britton, Kelly, & Cable, 2016; Palardy & Vermunt, 2010). Schools are therefore more likely to be classified into the same group with other schools that have similar distribution of individual-level latent-class patterns. Because NP-MLCA is a model-based technique using probability models, model-fit indices can be utilized to identify optimal solutions, in terms of the number of latent classes (i.e., mixture components). The probability of a person endorsing the response of the observed indicators can be estimated, and so can the latent class probability (also known as latent-class prevalence or mixture weights). Since all latent classes are identified simultaneously, the statistical power problems caused by reduced sample sizes can also be mitigated. Another advantage of NP-MLCA's model-based approach is that it provides researchers with the flexibility to incorporate covariates into the model to contextualize latent classes while identification of those classes is still ongoing (also known as a one-step procedure) (Bolck, Croon, & Hagenaars, 2004b). Classification error can thus be taken into account during the iteration process of estimation, when the relationship of interest is that between the covariates and the latent-class variables.

In the following of the paper, the next section describes the model specification of different mixture models. Then, Section 3 uses the empirical data from the Michigan 21st Century Community Learning Centers Grant Program (21st CCLC) to illustrate the conditional NP-MLCA. In Section 4, we use the empirical example as the basis for simulations that demonstrate the sensitivity of the model among 96 conditions. Section 5 concludes the study with a discussion of results.

## 2. Model Specification

### 2.1 Mixture Models with Covariates

One-step and three-step strategies have commonly been studied in mixture models, especially single-level latent class analysis (LCA). The covariate-conditioned posterior probability used in the one-step approach for identifying latent class membership is analogous to the multiple-indicator/multiple-cause model developed in factor analysis (Vermunt, 2010), while the main purpose of adopting three-step procedures (i.e., the classify-analyze method) for LCA (Clogg, 1995) is to contextualize the latent class. The first of these estimates covariate effects and identifies latent classes simultaneously using a single model. The second identifies the latent class first; treats the estimated latent class membership as the observed class membership; and then analyzes the relationships between the covariates and the latent class variable. Although the purposes for using them are different, previous studies have shown that the one-step strategy produces unbiased and efficient parameter estimates, whereas the three-step strategy results require bias correction for which various methods have been proposed (Zsuzsa Bakk, Tekle, & Vermunt, 2013; Bolck, Croon, & Hagenaars, 2004a). There have been abundant simulation studies investigating covariate-incorporating LCA in terms of the performance of the covariate effect as well as the performance of parameter estimations in different sample size scenarios (e.g., Bandeen-Roche, Miglioretti, Zeger, & Rathouz, 1997; Vermunt, 2010; Wurpts & Geiser, 2014), and the performance of two strategies in NP-MLCA has been compared in simulation studies (Park & Yu, 2018a).

## 2.2 Non-parametric Multilevel Latent Class Analysis

Recently, more simulation work has incorporated covariates in multilevel latent class analysis (MLCA) to examine the effects on the probability of latent class at the first level or the second level (Finch & French, 2013; Henry & Muthén, 2010b; Muthén & Asparouhov, 2009; Muthén & Asparouhov, 2014; Vermunt, 2003). Bennink, Croon, & Vermunt (2015) proposed a method to correct the bias of the relationship in the mediation model at the second-level when the mediator was a latent class variable measured using level-1 indicators. Using a parametric approach, Van Horn et al. (2008) examined the group-level intervention effects using a multilevel mixture model, and Van Horn et al. (2016) used multilevel regression mixture models to identify level-1 latent class in level-2 covariate effects. Using a nonparametric approach, Chang (2016) examined the required sample size of the NP-MLCA and conditional NP-MLCA. Similar results were also shown by (Park & Yu, 2018a), which compared strategies of incorporating covariates in the model under different degrees of class separation. The current study builds directly on the work of Chang (2016), extending this research and that of Park & Yu (2018a), by investigating the power and type-I error of the cross-level covariate estimation performance.

Let subject $i$ = 1, 2, …, $n_j$, and group $j$=1, 2, …, $J$. $\boldsymbol{Y_{ij}}$ is the response pattern to $K$ indicators of subject $i$ in group $j$, and $\boldsymbol{s}$ is a vector of a specific response pattern. $s_k$ denotes the response to the $k$-th indicator. $Y_{ijk}$ is the item response for subject $i$ in group $j$ to indicator $k$. $C_{ij}$ denotes the latent class variable at the individual level, and the individual-level latent class membership $c$ = 1, 2, …, $L$, where $L$ denotes the total number of the individual-level latent classes. Without the covariates in the NP-MLCA, the responses of the indicators in each latent class can be modeled by the product of the latent class probabilities, $P(C_{ij} = c)$, and the conditional response probabilities (CRP or the quality of indicators), $P(Y_{ijk} = s_k | C_{ij} = c)$.

Let $W_j$ denote the higher-level discrete latent class variable to which the $j$-th school belongs. The individual response pattern in MLCA model given their higher-level latent class membership $m = 1...M$, where $M$ denotes the total number of higher-level latent classes, can be specified as follows when local independence assumption holds:

$$P(Y_{ij} = s | W_j = m) = \sum_{c=1}^{L} P(C_{ij} = c | W_j = m) \prod_{k=1}^{K} P(Y_{ijk} = s_k | C_{ij} = c, W_j = m) \tag{1}$$

Multiplying the higher-level latent class probability, $P(W_j = m)$, we can know the probability of individual's response patterns is a function of the second-level latent class probability, the first-level latent class probability, and their CRP:

$$P(Y_{ij} = s) = \sum_{m=1}^{M} P(W_j = m) \times \left( \sum_{c=1}^{L} P(C_{ij} = c | W_j = m) \prod_{k=1}^{K} P(Y_{ijk} = s_k | C_{ij} = c, W_j = m) \right) \tag{2}$$

Since the level-2 latent class $W_j$, the level-1 latent class $C_{ij}$, and the responses of indicators $Y_{ijk}$ in the model are all discrete random variables, level-2 units the equation can be formulated to be more specific, using $h$ as the rolling index:

$$P(Y_{ij} = s) = \sum_{m=1}^{M} \frac{\exp(\alpha_m)}{\sum_{h=1}^{M} \exp(\alpha_h)} \sum_{c=1}^{L} \frac{\exp(\gamma_{cj})}{\sum_{h=1}^{L} \exp(\gamma_{hj})} \prod_{k=1}^{K} \frac{\exp(\beta_{s_k cj}^{k})}{\sum_{h=1}^{S_k} \exp(\beta_{hcj}^{k})} \tag{3}$$

The parameter $\alpha_m, \gamma_{cj}, \beta_{s_k cj}^{k}$ are described in the different contexts below.

Rather than restricting all the random intercepts $\gamma_{cj}$ to a scale continuum with a normal distribution as in the parametric approach, the non-parametric approach assumes the $\gamma_{cj}$ are discretely distributed. In other words, the random intercept is allowed to vary only across the level-2 latent classes instead of level-2 units, letting the higher-level units in the same mixture component (i.e., higher-level latent class membership) share the same parameter values. Thus, we can further replace $\gamma_{cj}$ with $\gamma_{cm}$ in NP-MLCA for model estimation. The intercept $\alpha_m$ denotes the log odds of a school being categorized to the level-2 latent class. In addition, $\beta^k_{s_k cj}$ denotes the log odds of a person in school $j$ has response $s_k$ given his/her latent class membership of class $c$. For simplicity, CRPs are assumed to be the same across schools (i.e., assuming invariant measurement error). We can simplify the model as follows:

$$P(Y_{ij} = s) = \sum_{m=1}^{M} \frac{\exp(\alpha_m)}{\sum_{h=1}^{M} \exp(\alpha_h)} \sum_{c=1}^{L} \frac{\exp(\gamma_{cm})}{\sum_{h=1}^{L} \exp(\gamma_{hj})} \prod_{k=1}^{K} \frac{\exp(\beta^k_{s_k c})}{\sum_{h=1}^{S_k} \exp(\beta^k_{hc})} \quad (4)$$

For model identification and effect coding, $\alpha_1$, one of the random intercepts, $\gamma_{c1}$, in the latent class probability, and one of the intercepts for the indicator $j$, $\beta^k_{1c}$, are set to 0 as the reference group. Also, since the indicators, the individual-level latent class and the school-level latent class are discrete random variables, $\sum_{s_k=1}^{S_k} P(Y_{ik} = s_k) = \sum_{c=1}^{L} P(C_i = c) = \sum_{m=1}^{M} P(W_j = m) = 1$.

**2.3 Non-parametric Multilevel Latent Class Analysis with Covariates**

There are several ways to incorporate covariates of interest in the NP-MLCA. The effects of level-1 level-2 covariates on the CRP (i.e., multilevel regression mixture modeling) on the first-level latent class probability and/or the second-level latent class probabilities can be studied for different research purposes. Van Horn et al. (2015) examined the group-level effects on the CRPs and Park & Yu (2018a) has simulation work on NP-MLCA models focusing on level-2 covariates effects on the level-2 latent classes and level-1 covariate effects on the level-1 latent classes.

In this study, to make the simulations closer to reality, we focused on the effect of level-1 and level-2 covariates on the level-1 latent class probability. In other words, the level-2 covariate effects are cross-level covariate effects. In addition, both nuisance effect (i.e., when no effect is true) and strong effect were considered in the conditions. In the condition of nuisance effect, the covariates were included but did not affect the students' probability of latent class membership at the level-1. We investigate whether adding nuisance covariates is helpful or detrimental to classification accuracy, and whether the model can recover the parameter precisely even if the effect is no different from zero. We also extended the scenario from the traditional two latent classes to three latent classes at level-1 with two consistent response patterns and one mixing response pattern in the level-1 latent classes. The model in this study was

$$P(Y_{ij} = s) = \sum_{m=1}^{M} \frac{\exp(\alpha_m)}{\sum_{h=1}^{M} \exp(\alpha_h)} \sum_{c=1}^{L} \frac{\exp(\gamma_{0cm} + \gamma_{1c}X_{ij} + \gamma_{2c}Z_j)}{\sum_{h=1}^{L} \exp(\gamma_{0cm} + \gamma_{1c}X_{ij} + \gamma_{2c}Z_j)} \prod_{k=1}^{K} \frac{\exp(\beta_{s_k c}^{k})}{\sum_{h=1}^{S_k} \exp(\beta_{hc}^{k})} \quad (5)$$

with $X_{ij}$ denoting the level-1 covariates, and $\gamma_{1c}$ denoting the effect of $X_{ij}$. $Z_j$ denotes the higher-level covariate, and $\gamma_{2c}$ denotes the cross-level effect (i.e., slope) of the covariate $Z_j$ on classifying to level-1 latent class probability. Since the outcome of interest is level-1 latent class, for each covariate, $(L-1)$ covariate effects are estimated. Figure 1 shows the conceptual model for the model in this study.

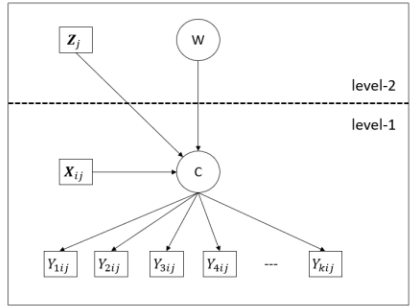

*Figure 1*. Conditional NP-MLCA

**2.4 Previous Simulation Studies**

Simulation studies in single-level LCA indicated that the one-step method was better when it came to parameter recovery, while the three-step method resulted in systematic errors in parameter estimation, requiring correction formulas for unbiased estimates (Bolck et al., 2004a). In NP-MLCA, very few papers have hitherto discussed how model parameter estimation can be performed when these two methods are applied, when covariates are incorporated, and in multilevel contexts. Finch and French (2013) indicated that the parameter recovery rate of the covariate effect and the quality of indicators (CRPs) were both good when the covariate was incorporated into NP-MLCA. However, their study fixed the association between the covariate and the latent class variable at one, and fixed the quality of indicators at [0.2, 0.8], providing us with no clues as to what differences might arise from varying covariate effects or varying indicator-quality levels. Park & Yu (2018a) compared the one-step and three-step strategies in NP-MLCA under various scenarios of class separation at both levels; however, cross-level covariate effects and nuisance covariate effects were not examined.

Mixed-pattern latent classes (i.e., indicators in the latent class have a mixed of high and low CRPs) has rarely been designed in previous simulation studies, but are commonly seen in latent class analyses in applied research (Cleveland, Collins, Lanza, Greenberg, & Feinberg, 2011;

Hirtenlehner, Starzer, & Weber, 2012; Mawditt et al., 2016). The degrees of mixing pattern in an identified latent class can vary depending on the number of indicators in the model. Previous literature demonstrated that high-quality indicators were beneficial to parameter estimation, and could compensate for the small effects of the covariates; however, the studies in question were focused on continuous latent variables (Marsh, Hau, Balla, & Grayson, 1998) or single-level LCA models (Wurpts & Geiser, 2014) and mixing patterns were not examined. Park & Yu (2018a, 2018b) considered the mixing patterns in the latent classes when suggesting the minimum samples sizes for NP-MLCA and compared the parameter estimation recovery of the covariates using one-step and three-step strategies. Their results indicated that the one-step method perform well in terms of parameter recovery and the accuracy of the standard error. But when the classical three-step method was utilized, the effects of level-1 covariates on the level-1 latent class probability were seriously downward-biased and the effects of level-2 covariates on the level-2 latent class probability were less biased. However, the study was unclear about the performance among specific latent classes.

This study therefore investigates the one-step method in a range of NP-MLCA contexts. In addition, for the sake of simplicity, $\alpha_m$ will be the same across the higher-level latent classes, i.e., in the model specified, the random component of the latent class probabilities only come from the level-1 latent class.

This study is organized into two sections. In the first section, we describe an empirical example using a multi-site program evaluation study. In the second section, we use the estimation of the empirical study to inform the design of the simulation study. The results of the simulation study are discussed at the end of the section to provide some insights of the model performance.

## 3. Empirical Example

### 3.1 Background

Program evaluation data from the Michigan 21st Century Community Learning Centers Grant Program (21st CCLC) were utilized to illustrate the model in this study. The program was created in 2002 to provide learning opportunities and activities for children in low-performing schools in Michigan by the state's Department of Education. It covers academic-enrichment and youth-development activities, drug and violence prevention programs, and art, music and recreation. Applicants for grants can be from school districts, public schools, universities, and nonprofit/community-based organizations. The funded grantees are evaluated yearly, and at the end of the program, teachers are asked to evaluate the students who participated as to whether they improved in terms of classroom behavior and homework performance.

### 3.2 Methods : Data Collection, Measures, and Covariates

For the purposes of this study, individual student improvement was measured in terms of the following 10 items: 1) Turning in homework on time, 2) Completing homework to your [the teacher's] satisfaction, 3) Participating in class, 4) Volunteering, 5) Attending class regularly, 6) Being attentive in class, 7) Behaving well in class, 8) Academic performance, 9) Coming to school motivated to learn, and 10) Getting along well with other students. If students had participated the program for more than 30 days, their teachers would be asked to mark the above items on a seven-point scale: significant decline, moderate decline, slight decline, no change, slight improvement, moderate improvement and significant improvement. For the sake of balanced sample size, the scale was dichotomized into improved (slight improvement and above) and unimproved (no change and below).

One of the study's focuses is to identify students in need of improving their classroom behavior and homework performance. In the case of students who took part in multiple programs, we randomly selected one program in which they participated for analysis. In addition, for estimation purposes, sites with less than five students were removed from the analysis. There were 29 sites (9.12% of total sites) removed due to small site sizes, which reduced the sample size from 6684 to 6580. The 10 latent class indicators and covariates were complete cases. Statistical software R (R Core Team, 2016) was used for data cleaning, and M*plus* 8.0 (Muthén & Muthén, 1998-2018) was utilized for model estimation.

The procedures above resulted in data from 6,580 individuals and 289 sites. The 10 evaluation items were used as the measured indicators for individual-level classification. The 289 sites at the higher-level units were classified using the model specified in equations (5) in section 2. Because this empirical example is intended to illustrate the proposed methodology, only a limited set of variables was included as the covariates in the model. Covariates explored at the individual level included sex, dummy-coded ethnicity, and activity participation hours. Covariates explored at the site (i.e., grantee) level included number of program staff.

### 3.3 Results

Table 1 shows the CRPs of individuals endorsing the "improved" answer given their latent class membership in the conditional NP-MLCA model. The main role of the CRPs is to characterize the individual latent classes. As shown in the table, individuals in Improved Group endorsed highly (> 0.93) on the "Improved" category. In contrast, individuals classified in Non-improved Group endorsed low (0.15 and less) on the "Improved" category. The third class, referred to "Mixed Group", endorsed moderately across these ten indicators. If we look closely at the probability pattern on these ten items in this group, we can find that for this group of students, their teachers reported that the individuals' school behaviors were improved, but that they struggled more with the academic work.

Table 1

*Conditional Response Probabilities from the Conditional NP-MLCA model*

| Conditional Response Probabilities | Conditional NP-MLCA (N=6580, J = 289) | | |
|---|---|---|---|
| | Improved Group (n = 2301, 34.97%) | Mixed Group (n = 1783, 27.1%) | Non-improved Group (n = 2496, 37.93%) |
| HW on time | 0.932 | 0.389 | 0.044 |
| HW satisfaction | 0.945 | 0.353 | 0.026 |
| Class Participation | 0.933 | 0.344 | 0.024 |
| Volunteering | 0.969 | 0.664 | 0.126 |
| Class Attendance | 0.983 | 0.732 | 0.153 |
| Attentive in Class | 0.989 | 0.533 | 0.008 |
| Class Behavior | 0.979 | 0.637 | 0.024 |
| Academic Performance | 0.944 | 0.305 | 0.005 |
| Motivated to Learn | 0.993 | 0.525 | 0.008 |
| Getting Along with Other Students | 0.974 | 0.589 | 0.038 |

Table 2

*Model Fit Information and Classification Quality*

| | Conditional NP-MLCA |
|---|---|
| Number of free parameters | 45 |
| Entropy | 0.896 |
| AIC | 51116.890 |
| BIC | 51422.521 |

Two types of sites were found in this study, referred to Type I and Type II sites. 75.38% of the sites are Type I while 24.62% of the sites are type II. The composition of site-level latent classes in the NP-MLCA model is shown in Figures 3. The latent class membership estimated from this model has controlled for the demographic variables at the individual level and the resources at the site level. Figure 3 shows that two types of sites have different compositions of the student-level latent class. Sites that were categorized as Type I have larger proportion of students in the improved group. In Type I, sites have larger proportion of improved students (43.34%), followed by 30.67% of the students are in mixed group, and 29.34% of the students in the site was in non-improved group. In contrast, sites in Type II had larger proportion in the non-improved group (62.82%), followed by 17.96% of the students in the site were mixed group, and 12.76% of the students were improved group. The latent class variables at the student level and the site/grantee level and the contextual effect and cross-level effect of the predictors were estimated at the same time.

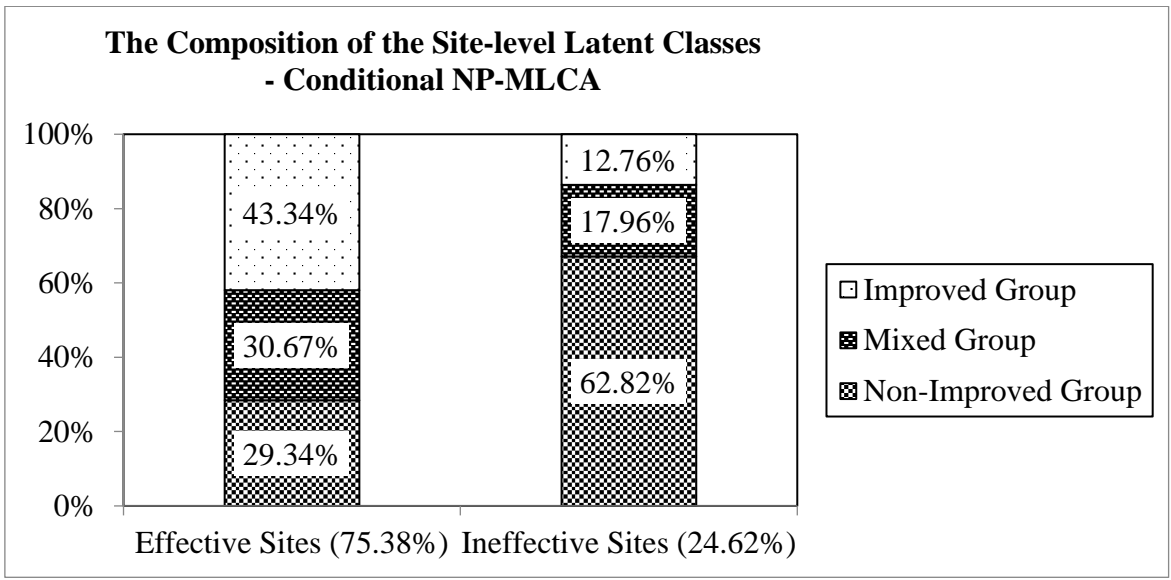

*Figure 3.* The composition of site-level latent classes in conditional NP-MLCA model.

Table 3

*The Level-1 Covariate Effects and Level-2 Cross-level Effect on Level-1 Latent Class Solution (Scale: Odds Ratio)*

| Predictors | Reference Group | Comparison of non-improved group to improved group | | Comparison of mixed group to improved group | |
|---|---|---|---|---|---|
| | | Estimate | CI | Estimate | CI |
| *Level - 1 Predictors* | | | | | |
| Female | Male | 0.785** | (0.686-0.899) | 0.852* | (0.740-0.981) |
| White | Non-White | 0.903 | (0.731-1.116) | 1.263* | (1.042-1.530) |
| Special Education | No | 1.011 | (0.653-1.565) | 1.732* | (1.193-2.513) |
| Non-Academic Day |  | 1.010*** | (1.006-1.014) | 1.007** | (1.003-1.011) |
| *Level - 2 Predictor* |  |  |  |  |  |
| Number of Staffs |  | 0.950* | (0.908-0.994) | 0.927*** | (0.888-0.968) |

*Note: * p < .05, ** p < .01, *** p < .001*

According to the results shown in NP-MLCA with covariates in Table 3, four student-level contextual effects: female, white, special education, and non-academic day, and one site-level cross-level effect: number of staffs were statistically significant in affecting students' latent class probability. Males are more likely to be in the improved group, as opposed to the non-improved and mixed groups. For males, the odds of being categorized in the improved group are 1.274 (1/0.785) times more likely than the odds for a female categorized in an improved group. The odds of white students to be categorized in a mixed group are 1.263 times more likely than the odds of a non-white student. Special education students are 73% more likely to be in the mixed group rather than in the improved group. For students enrolled for one additional non-academic day, we would see about 1% increase in the odds of being categorized into the mixed group or the non-improved group. For students having one additional staff at the site, they have 1.053 times more likely to be in the improved group rather than in the non-improved group, and 1.079 times more likely to be in the improved group rather than in the mixed group. In other words, students enrolled in the site with more staffs increase their probabilities improve.

## 4. Simulation Study

### 4.1 Purposes

Although NP-MLCA has been widely applied in multiple disciplines, two major issues confront its use in multilevel contexts: how to accurately classify individual-level and group-level units, and how to provide precise model parameter estimations when covariates are included in the model. Therefore, the study intends to answer how the quality of two outcome measures – parameter estimation and classification accuracies – are affected by the following five study factors at both levels in NP-MLCA: 1) the quality of latent class indicators (i.e., conditional response parameters), 2) the number of latent class indicators, 3) the effect size of covariates at different levels, 4) the number of groups/higher-level units (e.g., programs, schools, hospitals, companies), and 5) the size of groups/higher-level units (e.g., the number of students within each school, the number of patients within each hospital, and the number of personnel within each company). A total of 96 conditions were investigated.

## 4.2 Methods

*4.2.1 Simulation Design*

Mimicking the scenario of the multi-site program evaluation case in the empirical study, the number of individual-level latent classes is fixed at 3, and the number of site-level latent classes is fixed at 2 in the simulation study. All indicators were binary variables. The separation of the latent classes in this study was fixed at $\log(2.5)$ for Class 1 and $\log(1.5)$ for Class 2, with Class 3 as the reference class. Different levels of six study factors were varied. 1) The quality of latent class indicators (i.e., CRPs) was varied in three conditions: [0.1, 0.9], [0.2, 0.8], and [0.3, 0.7] (i.e., individuals with high CRPs for all of the indicators are in Class 1; individuals with high CRPs for half of the indicators, and low CRPs for the other half of the indicators, are in Class 2; individuals with low CRPs for all of the indicators are in Class 3; in other words, individuals in Classes 1 and 3 have consistent response patterns across all indicators, while individuals in Class 2 have mixed response patterns). 2) the number of latent class indicators: 6 and 12. 3) the number of sites: 50 and 150. 4) The site size: 30 and 60. 5) The covariate effects in odds ratio unit at the individual level: (1, 1), (1.5, 3). 6) The covariate effects in odds ratio unit at the site level: (1, 1), (1.5, 3). The pair of the covariate effects denotes the effect of the covariate on the first latent class and the second latent class compared to the reference class in the scale of odds ratio. For example, if the covariate coded females as 1, males as 0, the covariate effects (1.5, 3) at the student level mean that the odds of a female being categorized in the improved group are 1.5 times greater than the odds for a male categorized in an improved group, and the odds of being categorized in the mixed group are 3 times greater than the odds for a male categorized in a mixed group.

The combination of the six study factors resulted in 96 conditions. For each of the conditions, 500 replication datasets were simulated. R (R-3.4.1, R Core Team, 2016) was used to generate the data and extract and calculate the parameter estimates recovery and latent class classification accuracy after modeling. The statistical software M*plus* 7.11 (Muthén & Muthén, 2018) was used to fit the model to the generated data.

*4.2.2 Evaluation Criteria*

The recovery of parameter estimation and the classification accuracy were considered the outcome measures and examined if the six study factors listed above have significant effects on these outcome measures.

Specifically, classification accuracy was measured via proportional consistency between true latent class membership and the predicted latent class membership at each level. The quality of parameter estimation of CRPs was evaluated in terms of bias, standard error estimate, and the ratio of standard error estimate to standard deviation. The bias is obtained by taking the difference between the average estimate and the population parameter. The quality of the standard error estimates is evaluated by how close they are to the standard deviation of the parameter estimates. The standard error to standard deviation ratio was calculated by dividing the average of the standard error of the estimates by the standard deviation of the parameter estimates. The standard error is underestimated if the ratio is less than 1 and overestimated otherwise. For the quality of the covariate estimates, we evaluated the Type I error for the nuisance variables and the power for the variables with effects.

Because latent classes at both levels were categorical, label switching problems can happen at both levels simultaneously. There were three latent classes at the first level and two latent classes at the second level. This design can result in $2^3 = 8$ possible switching combinations in the measurement model at level-1 and two switching combinations at level-2. We first identified which label switching case the replication output was, based on the CRP, then applied logarithm properties to recalculate the parameter estimates so that every replication was in the same order of class labels. We further aggregated the level-1 latent class probability to the higher level to evaluate if the level-2 labels had switched or not. After the switching issues were sorted out and the switching cases were identified, the covariate estimates were recalculated accordingly. Since there is no easy solution to relabel the standard error of the estimate in the three latent classes scenario, we only used the non-switched cases of the replication datasets, comparing with the standard deviation of parameter from the corresponding datasets to evaluate the quality of the standard error estimates. Additional details of simulation procedure and label switching issues are addressed in Chang (2019) and M*plus* code for the conditional NP-MLCA in this paper is available from the authors on request.

## 4.3 Simulation Results

To evaluate the model performance, results of 96 conditions were examined in terms of parameter recovery and classification accuracy.

*4.3.1 Parameter Recovery*

*Conditional Response Probability.* The bias, standard error, standard deviation, and the ratio of the standard error to the standard deviation of the measurement model parameter (i.e., CRP) and the covariate effects at two levels across all 96 conditions are presented in Table 4.

Overall, most of the bias of CRP in the conditions were less than 0.01 and the standard error/standard deviation ratios were very close to 1, except for the conditions where the number of indicators were 6 and nuisance covariate effects were included at both levels (i.e., the condition where the covariates were included but had no effect (1, 1) or no relationships with the level-1 latent class). The bias reached above 0.01 when the conditions were 0.7 CRP, 50 sites, and 30 individuals per site. Despite the higher bias found in these conditions, the result is within our expectation since there was no distinct separation in latent classes, and incorporated covariates provided no additional information. When non-nuisance effects in covariates (i.e., conditions with (1.5, 3) covariate effects) were included in the model, differences across latent classes in the amount of bias of CRP became trivial. From Table 4, we can see that increasing the sample size or including large-effect covariates substantially decreased the bias of the CRP estimate, bringing it down to the third decimal place.

The results were similar for standard error estimates. The standard errors were underestimated for the conditions in which the number of indicators was 6, the CRP was 0.7, the number of sites was 50, the site size was 30, and the model included nuisance covariates. The ratio is as low as 0.523 for latent class 2 and 0.628 for latent class 3. The reasons for these low ratios are similar to those for the bias of the CRP estimates: when the class separation is small (i.e., the quality of the indicators is low) and the sample size is small, including nuisance covariates can be detrimental to recovering the parameters in the measurement model. On the contrary, including non-nuisance covariates in the models when the sample size is small can improve the CRP estimates (as seen in the last columns in Table 4). The standard errors were well-recovered across the conditions in the small sample size scenarios.

Overall, when the sample size was small, increasing the quality of CRPs to [0.2, 0.8], increasing the number of indicators to 12 items, or including strong-effect covariates in the model helped decrease the bias of the parameter estimates and improve the accuracy of the standard error estimates.

Table 4

*Conditional Response Probability Estimate Recovery of Latent Classes*

| Number of Indicators | Quality of Indicators | Number of Sites | Site Size | \multicolumn{4}{c}{Level-1 Level-2 Covariate Effects (1, 1)(1, 1)} | \multicolumn{4}{c}{(1, 1)(1.5, 3)} | \multicolumn{4}{c}{(1.5, 3)(1, 1)} | \multicolumn{4}{c}{(1.5, 3)(1.5, 3)} |
|---|---|---|---|---|---|---|---|---|---|---|---|---|---|---|---|---|---|---|---|
| | | | | bias | SE | SD | SE/SD | bias | SE | SD | SE/SD | bias | SE | SD | SE/SD | bias | SE | SD | SE/SD |
| \multicolumn{20}{c}{Latent Class 1} |
| 12 | 0.7 | 50 | 60 | 0.001 | 0.083 | 0.084 | 0.991 | 0.001 | 0.085 | 0.086 | 0.986 | 0.002 | 0.084 | 0.086 | 0.973 | 0.001 | 0.085 | 0.087 | 0.984 |
| 12 | 0.7 | 150 | 30 | 0.001 | 0.069 | 0.070 | 0.987 | 0.001 | 0.070 | 0.070 | 0.988 | 0.001 | 0.070 | 0.069 | 1.012 | 0.002 | 0.070 | 0.070 | 0.993 |
| 12 | 0.7 | 150 | 60 | 0.002 | 0.048 | 0.049 | 0.979 | 0.001 | 0.049 | 0.050 | 0.989 | 0.000 | 0.049 | 0.049 | 1.002 | 0.000 | 0.049 | 0.050 | 0.979 |
| 12 | 0.8 | 50 | 30 | 0.005 | 0.110 | 0.110 | 1.003 | 0.008 | 0.113 | 0.116 | 0.978 | 0.005 | 0.112 | 0.114 | 0.987 | 0.007 | 0.116 | 0.118 | 0.978 |
| 12 | 0.8 | 50 | 60 | 0.001 | 0.077 | 0.078 | 0.986 | 0.000 | 0.080 | 0.080 | 0.991 | 0.001 | 0.080 | 0.080 | 0.993 | 0.000 | 0.081 | 0.083 | 0.976 |
| 12 | 0.8 | 150 | 30 | 0.001 | 0.064 | 0.065 | 0.989 | 0.000 | 0.066 | 0.068 | 0.967 | 0.001 | 0.065 | 0.066 | 0.987 | 0.000 | 0.067 | 0.067 | 1.007 |
| 12 | 0.8 | 150 | 60 | 0.001 | 0.045 | 0.046 | 0.990 | 0.000 | 0.046 | 0.047 | 0.991 | 0.000 | 0.046 | 0.047 | 0.993 | 0.000 | 0.048 | 0.048 | 0.999 |
| 12 | 0.9 | 50 | 30 | 0.011 | 0.137 | 0.138 | 0.989 | 0.013 | 0.141 | 0.143 | 0.990 | 0.012 | 0.142 | 0.143 | 0.992 | 0.012 | 0.144 | 0.148 | 0.972 |
| 12 | 0.9 | 50 | 60 | 0.005 | 0.096 | 0.097 | 0.992 | 0.004 | 0.099 | 0.101 | 0.988 | 0.004 | 0.099 | 0.101 | 0.974 | 0.003 | 0.102 | 0.105 | 0.975 |
| 12 | 0.9 | 150 | 30 | 0.002 | 0.079 | 0.081 | 0.983 | -0.001 | 0.082 | 0.083 | 0.981 | 0.001 | 0.082 | 0.082 | 1.003 | 0.000 | 0.084 | 0.085 | 0.986 |
| 12 | 0.9 | 150 | 60 | 0.001 | 0.056 | 0.057 | 0.991 | 0.002 | 0.058 | 0.058 | 1.004 | 0.002 | 0.058 | 0.059 | 0.985 | 0.001 | 0.060 | 0.060 | 0.996 |
| \multicolumn{20}{c}{Latent Class 2} |
| 12 | 0.7 | 50 | 30 | -0.002 | 0.148 | 0.145 | 1.024 | 0.001 | 0.127 | 0.126 | 1.008 | 0.000 | 0.130 | 0.128 | 1.020 | 0.000 | 0.120 | 0.119 | 1.010 |
| 12 | 0.7 | 50 | 60 | -0.001 | 0.096 | 0.098 | 0.982 | 0.000 | 0.086 | 0.088 | 0.986 | 0.001 | 0.086 | 0.088 | 0.983 | -0.001 | 0.084 | 0.083 | 1.006 |
| 12 | 0.7 | 150 | 30 | -0.001 | 0.081 | 0.079 | 1.024 | -0.001 | 0.071 | 0.070 | 1.014 | 0.002 | 0.071 | 0.072 | 0.989 | -0.001 | 0.067 | 0.067 | 0.995 |
| 12 | 0.7 | 150 | 60 | 0.002 | 0.056 | 0.056 | 0.987 | 0.001 | 0.050 | 0.050 | 1.004 | 0.001 | 0.050 | 0.050 | 1.006 | 0.001 | 0.047 | 0.048 | 0.989 |
| 12 | 0.8 | 50 | 30 | 0.001 | 0.116 | 0.119 | 0.978 | 0.000 | 0.113 | 0.114 | 0.987 | -0.001 | 0.114 | 0.115 | 0.984 | 0.000 | 0.111 | 0.114 | 0.973 |
| 12 | 0.8 | 50 | 60 | -0.001 | 0.081 | 0.084 | 0.966 | 0.000 | 0.079 | 0.080 | 0.988 | 0.000 | 0.080 | 0.080 | 0.992 | -0.001 | 0.078 | 0.078 | 0.994 |
| 12 | 0.8 | 150 | 30 | -0.001 | 0.067 | 0.067 | 1.005 | 0.001 | 0.065 | 0.066 | 0.986 | 0.001 | 0.065 | 0.066 | 0.993 | -0.001 | 0.064 | 0.064 | 1.005 |
| 12 | 0.8 | 150 | 60 | 0.001 | 0.047 | 0.047 | 0.998 | 0.000 | 0.046 | 0.046 | 1.007 | 0.000 | 0.046 | 0.046 | 1.018 | 0.000 | 0.045 | 0.046 | 0.983 |
| 12 | 0.9 | 50 | 30 | 0.000 | 0.139 | 0.141 | 0.980 | 0.000 | 0.136 | 0.139 | 0.976 | -0.001 | 0.137 | 0.141 | 0.971 | 0.001 | 0.137 | 0.141 | 0.970 |
| 12 | 0.9 | 50 | 60 | 0.001 | 0.097 | 0.100 | 0.973 | -0.001 | 0.096 | 0.099 | 0.971 | 0.001 | 0.096 | 0.098 | 0.980 | -0.001 | 0.096 | 0.097 | 0.983 |
| 12 | 0.9 | 150 | 30 | -0.001 | 0.080 | 0.080 | 1.009 | 0.000 | 0.079 | 0.079 | 1.002 | 0.000 | 0.079 | 0.080 | 0.994 | -0.002 | 0.078 | 0.078 | 1.000 |
| 12 | 0.9 | 150 | 60 | 0.000 | 0.057 | 0.057 | 0.991 | 0.000 | 0.056 | 0.056 | 0.990 | 0.000 | 0.056 | 0.057 | 0.995 | -0.001 | 0.055 | 0.056 | 0.985 |
| \multicolumn{20}{c}{Latent Class 3} |
| 12 | 0.7 | 50 | 30 | -0.001 | 0.197 | 0.197 | 1.001 | -0.008 | 0.170 | 0.170 | 1.000 | -0.004 | 0.165 | 0.165 | 0.998 | -0.003 | 0.153 | 0.155 | 0.988 |
| 12 | 0.7 | 50 | 60 | -0.007 | 0.136 | 0.135 | 1.014 | -0.003 | 0.119 | 0.117 | 1.012 | -0.002 | 0.117 | 0.115 | 1.025 | -0.001 | 0.106 | 0.106 | 0.997 |
| 12 | 0.7 | 150 | 30 | -0.004 | 0.109 | 0.110 | 0.989 | -0.001 | 0.096 | 0.094 | 1.021 | -0.001 | 0.095 | 0.095 | 1.008 | -0.004 | 0.089 | 0.088 | 1.004 |
| 12 | 0.7 | 150 | 60 | 0.001 | 0.076 | 0.076 | 0.998 | 0.001 | 0.067 | 0.067 | 0.996 | 0.001 | 0.067 | 0.068 | 0.980 | -0.001 | 0.062 | 0.062 | 1.006 |
| 12 | 0.8 | 50 | 30 | -0.008 | 0.163 | 0.167 | 0.978 | -0.007 | 0.154 | 0.157 | 0.983 | -0.005 | 0.151 | 0.152 | 0.996 | -0.005 | 0.144 | 0.149 | 0.965 |
| 12 | 0.8 | 50 | 60 | -0.006 | 0.115 | 0.114 | 1.004 | -0.004 | 0.107 | 0.110 | 0.978 | -0.003 | 0.108 | 0.108 | 0.999 | -0.002 | 0.102 | 0.104 | 0.975 |
| 12 | 0.8 | 150 | 30 | -0.004 | 0.094 | 0.096 | 0.969 | -0.001 | 0.088 | 0.087 | 1.011 | -0.002 | 0.088 | 0.088 | 1.000 | -0.003 | 0.085 | 0.084 | 1.007 |
| 12 | 0.8 | 150 | 60 | -0.001 | 0.066 | 0.065 | 1.020 | 0.000 | 0.062 | 0.063 | 0.988 | 0.000 | 0.062 | 0.062 | 0.998 | -0.001 | 0.060 | 0.060 | 0.997 |
| 12 | 0.9 | 50 | 30 | -0.014 | 0.195 | 0.204 | 0.957 | -0.011 | 0.188 | 0.193 | 0.971 | -0.011 | 0.188 | 0.192 | 0.979 | -0.008 | 0.181 | 0.185 | 0.977 |
| 12 | 0.9 | 50 | 60 | -0.007 | 0.137 | 0.139 | 0.991 | -0.005 | 0.132 | 0.138 | 0.961 | -0.006 | 0.132 | 0.134 | 0.986 | -0.004 | 0.126 | 0.132 | 0.961 |
| 12 | 0.9 | 150 | 30 | -0.006 | 0.114 | 0.116 | 0.978 | -0.004 | 0.109 | 0.109 | 0.997 | -0.007 | 0.109 | 0.110 | 0.996 | -0.004 | 0.106 | 0.106 | 0.997 |
| 12 | 0.9 | 150 | 60 | -0.003 | 0.080 | 0.080 | 1.001 | -0.001 | 0.077 | 0.078 | 0.984 | -0.001 | 0.077 | 0.077 | 1.001 | -0.001 | 0.075 | 0.075 | 1.000 |

| Level-1 | Level-2 | Covariate Effects | | (1, 1)(1, 1) | | | | (1, 1)(1.5, 3) | | | | (1.5, 3)(1, 1) | | | | (1.5, 3)(1.5, 3) | | | |
|---|---|---|---|---|---|---|---|---|---|---|---|---|---|---|---|---|---|---|---|
| Number of Indicators | Quality of Indicators | Number of Sites | Site Size | bias | SE | SD | SE/SD | bias | SE | SD | SE/SD | bias | SE | SD | SE/SD | bias | SE | SD | SE/SD |
| | | | | Latent Class 1 | | | | | | | | | | | | | | | |
| 6 | 0.7 | 50 | 30 | 0.028 | 0.294 | 0.283 | 1.036 | 0.018 | 0.206 | 0.205 | 1.003 | 0.020 | 0.210 | 0.199 | 1.059 | 0.012 | 0.210 | 0.189 | 1.110 |
| 6 | 0.7 | 50 | 60 | 0.016 | 0.163 | 0.149 | 1.090 | 0.009 | 0.140 | 0.132 | 1.058 | 0.009 | 0.131 | 0.129 | 1.010 | 0.004 | 0.123 | 0.129 | 0.952 |
| 6 | 0.7 | 150 | 30 | 0.010 | 0.129 | 0.123 | 1.052 | 0.008 | 0.107 | 0.105 | 1.026 | 0.009 | 0.110 | 0.108 | 1.017 | 0.003 | 0.108 | 0.102 | 1.055 |
| 6 | 0.7 | 150 | 60 | 0.002 | 0.082 | 0.081 | 1.011 | 0.004 | 0.072 | 0.074 | 0.971 | 0.005 | 0.073 | 0.072 | 1.012 | 0.003 | 0.072 | 0.073 | 0.993 |
| 6 | 0.8 | 50 | 30 | 0.008 | 0.150 | 0.148 | 1.011 | 0.007 | 0.148 | 0.151 | 0.985 | 0.008 | 0.152 | 0.147 | 1.029 | 0.010 | 0.149 | 0.148 | 1.008 |
| 6 | 0.8 | 50 | 60 | 0.004 | 0.100 | 0.103 | 0.969 | 0.007 | 0.102 | 0.103 | 0.988 | 0.003 | 0.102 | 0.102 | 0.992 | 0.004 | 0.103 | 0.106 | 0.974 |
| 6 | 0.8 | 150 | 30 | 0.005 | 0.084 | 0.084 | 1.000 | 0.004 | 0.084 | 0.083 | 1.012 | 0.002 | 0.085 | 0.082 | 1.032 | -0.001 | 0.084 | 0.084 | 1.007 |
| 6 | 0.8 | 150 | 60 | 0.001 | 0.059 | 0.058 | 1.018 | 0.003 | 0.059 | 0.059 | 0.998 | 0.003 | 0.059 | 0.058 | 1.019 | 0.000 | 0.059 | 0.059 | 1.009 |
| 6 | 0.9 | 50 | 30 | 0.007 | 0.152 | 0.157 | 0.970 | 0.008 | 0.157 | 0.160 | 0.983 | 0.003 | 0.156 | 0.156 | 0.999 | 0.008 | 0.162 | 0.163 | 0.994 |
| 6 | 0.9 | 50 | 60 | 0.004 | 0.107 | 0.110 | 0.971 | 0.004 | 0.111 | 0.112 | 0.989 | 0.005 | 0.111 | 0.113 | 0.983 | 0.005 | 0.113 | 0.112 | 1.006 |
| 6 | 0.9 | 150 | 30 | 0.004 | 0.088 | 0.086 | 1.019 | 0.004 | 0.091 | 0.093 | 0.983 | 0.005 | 0.091 | 0.094 | 0.970 | -0.001 | 0.093 | 0.093 | 0.997 |
| 6 | 0.9 | 150 | 60 | 0.001 | 0.062 | 0.063 | 0.990 | 0.000 | 0.064 | 0.064 | 0.991 | 0.002 | 0.064 | 0.065 | 0.993 | 0.001 | 0.066 | 0.064 | 1.025 |
| | | | | Latent Class 2 | | | | | | | | | | | | | | | |
| 6 | 0.7 | 50 | 30 | 0.037 | 0.508 | 0.808 | 0.628 | -0.017 | 0.227 | 0.258 | 0.879 | 0.000 | 0.227 | 0.216 | 1.049 | -0.004 | 0.176 | 0.176 | 1.001 |
| 6 | 0.7 | 50 | 60 | 0.014 | 0.214 | 0.242 | 0.886 | 0.000 | 0.142 | 0.138 | 1.026 | 0.003 | 0.132 | 0.134 | 0.986 | -0.003 | 0.126 | 0.117 | 1.080 |
| 6 | 0.7 | 150 | 30 | 0.006 | 0.198 | 0.190 | 1.045 | -0.002 | 0.108 | 0.108 | 1.007 | -0.003 | 0.109 | 0.107 | 1.025 | -0.002 | 0.091 | 0.093 | 0.975 |
| 6 | 0.7 | 150 | 60 | 0.004 | 0.111 | 0.113 | 0.980 | 0.000 | 0.073 | 0.074 | 0.994 | 0.003 | 0.072 | 0.072 | 1.002 | -0.001 | 0.062 | 0.064 | 0.984 |
| 6 | 0.8 | 50 | 30 | 0.003 | 0.183 | 0.185 | 0.988 | -0.004 | 0.155 | 0.158 | 0.977 | -0.002 | 0.162 | 0.158 | 1.023 | 0.000 | 0.143 | 0.146 | 0.978 |
| 6 | 0.8 | 50 | 60 | 0.002 | 0.126 | 0.127 | 0.987 | 0.001 | 0.105 | 0.108 | 0.977 | 0.002 | 0.106 | 0.109 | 0.975 | -0.001 | 0.098 | 0.104 | 0.939 |
| 6 | 0.8 | 150 | 30 | -0.001 | 0.102 | 0.101 | 1.011 | 0.000 | 0.089 | 0.088 | 1.009 | -0.005 | 0.090 | 0.087 | 1.025 | -0.001 | 0.082 | 0.084 | 0.971 |
| 6 | 0.8 | 150 | 60 | 0.000 | 0.070 | 0.070 | 1.004 | 0.000 | 0.062 | 0.062 | 0.994 | 0.004 | 0.062 | 0.060 | 1.041 | -0.002 | 0.057 | 0.057 | 0.994 |
| 6 | 0.9 | 50 | 30 | -0.003 | 0.167 | 0.171 | 0.981 | -0.002 | 0.161 | 0.165 | 0.973 | -0.004 | 0.158 | 0.163 | 0.969 | 0.000 | 0.154 | 0.154 | 1.001 |
| 6 | 0.9 | 50 | 60 | 0.002 | 0.116 | 0.122 | 0.957 | 0.002 | 0.112 | 0.113 | 0.989 | 0.000 | 0.111 | 0.112 | 0.997 | 0.001 | 0.108 | 0.113 | 0.955 |
| 6 | 0.9 | 150 | 30 | -0.002 | 0.096 | 0.095 | 1.008 | 0.000 | 0.092 | 0.094 | 0.976 | -0.003 | 0.093 | 0.092 | 1.009 | 0.000 | 0.090 | 0.092 | 0.981 |
| 6 | 0.9 | 150 | 60 | -0.001 | 0.067 | 0.068 | 0.988 | 0.001 | 0.065 | 0.065 | 0.996 | 0.004 | 0.065 | 0.064 | 1.006 | -0.002 | 0.063 | 0.064 | 0.989 |
| | | | | Latent Class 3 | | | | | | | | | | | | | | | |
| 6 | 0.7 | 50 | 30 | -0.236 | 0.542 | 1.036 | 0.523 | -0.026 | 0.264 | 0.276 | 0.955 | -0.017 | 0.290 | 0.260 | 1.116 | -0.015 | 0.218 | 0.220 | 0.991 |
| 6 | 0.7 | 50 | 60 | -0.068 | 0.303 | 0.445 | 0.680 | -0.013 | 0.168 | 0.179 | 0.938 | -0.010 | 0.178 | 0.175 | 1.021 | -0.013 | 0.149 | 0.152 | 0.983 |
| 6 | 0.7 | 150 | 30 | -0.034 | 0.242 | 0.258 | 0.937 | -0.008 | 0.144 | 0.141 | 1.020 | -0.003 | 0.143 | 0.139 | 1.026 | -0.003 | 0.120 | 0.121 | 0.989 |
| 6 | 0.7 | 150 | 60 | -0.011 | 0.152 | 0.149 | 1.025 | 0.000 | 0.094 | 0.097 | 0.965 | -0.004 | 0.097 | 0.100 | 0.975 | 0.000 | 0.084 | 0.085 | 0.988 |
| 6 | 0.8 | 50 | 30 | -0.026 | 0.242 | 0.260 | 0.930 | -0.015 | 0.200 | 0.206 | 0.972 | -0.003 | 0.201 | 0.205 | 0.977 | -0.005 | 0.184 | 0.184 | 0.999 |
| 6 | 0.8 | 50 | 60 | -0.015 | 0.169 | 0.168 | 1.009 | -0.007 | 0.140 | 0.143 | 0.978 | -0.009 | 0.138 | 0.141 | 0.979 | -0.007 | 0.128 | 0.128 | 1.002 |
| 6 | 0.8 | 150 | 30 | -0.006 | 0.135 | 0.134 | 1.007 | -0.004 | 0.114 | 0.117 | 0.971 | -0.003 | 0.113 | 0.111 | 1.017 | -0.001 | 0.106 | 0.104 | 1.015 |
| 6 | 0.8 | 150 | 60 | -0.007 | 0.096 | 0.096 | 0.999 | -0.002 | 0.080 | 0.082 | 0.976 | -0.003 | 0.079 | 0.081 | 0.974 | 0.000 | 0.074 | 0.074 | 0.999 |
| 6 | 0.9 | 50 | 30 | -0.026 | 0.230 | 0.235 | 0.978 | -0.016 | 0.216 | 0.224 | 0.966 | -0.006 | 0.216 | 0.219 | 0.987 | -0.014 | 0.203 | 0.209 | 0.975 |
| 6 | 0.9 | 50 | 60 | -0.011 | 0.160 | 0.164 | 0.973 | -0.007 | 0.148 | 0.151 | 0.984 | -0.009 | 0.151 | 0.152 | 0.992 | -0.011 | 0.145 | 0.148 | 0.975 |

*Covariate Estimates.* We evaluated the bias, Type-I error, and power of the covariate effects (shown in Table 5). The covariate estimates' bias and power were evaluated when the effect was designed to be 1.5 or 3 in logit scale. When the effect was designed to be 1 (i.e., a nuisance covariate), the estimates' biases and Type-I errors were evaluated.

Overall, the parameter recovery for the covariate estimates performed very well. Most of the biases were as low as the third decimal place except again for the condition with 6 indicators, 0.7 CRP, 50 sites, and 30 individuals per site, in which case biases were at the second decimal place. Another result that is worth pointing out is that when covariate effects were present at both levels, the effects on latent class 2 tended to be biased upward a little in those conditions with 6 indicators and 0.7 CRP. This may have been due to the mixed-pattern response of latent class 2. These biases were only as low as the second decimal place. These results were aligned with previous studies (Park & Yu, 2018a; Wurpts & Geiser, 2014). Researchers can avoid this by simply increasing the quality of indicators and/or the number of indicators.

Most of the Type I errors of the nuisance covariates were around 0.05, except for those nuisance effects on latent class 2 (i.e., CW2 on X and CW2 on W columns in Table 5). No matter whether the covariate effect was a first-level covariate effect or a cross-level covariate effect, the Type-I errors of the nuisance covariates' estimates were inflated. The inflation problems were alleviated by increasing the number of indicators to 12, although they still showed mild inflation in that case. Since latent class 2 is a mixed-patterned class, misidentifying the noise effect in the model is expected, and worth noting. We will discuss this further in the next section.

For those conditions for which the covariate effects were at (1.5, 3), the power mostly exceeded 0.8 except for, again, the condition that the number of indicators was 6, the CRP was 0.7, the number of sites was 50, and the site size was 30. Under this small sample size and low classification separation condition, no matter which level in which the covariates were incorporated, the power for the strong covariate effect reached 0.8 or higher, while the weak covariate effects were under-powered.

Table 5

*Bias and Power/ Type I Error of Covariate Effect at level-1 and level-2*

| Level-1 Level-2 Covariate Effects | | | | (1, 1)(1, 1) | | | | | | | | (1, 1)(1.5, 3) | | | | | | | | (1.5, 3)(1, 1) | | | | | | | | (1.5, 3)(1.5, 3) | | | | | | | |
|---|---|---|---|---|---|---|---|---|---|---|---|---|---|---|---|---|---|---|---|---|---|---|---|---|---|---|---|---|---|---|---|---|---|---|
| Parameter | | | | CW1 on X | | CW2 on X | | CW1 on W | | CW2 on W | | CW1 on X | | CW2 on X | | CW1 on W | | CW2 on W | | CW1 on X | | CW2 on X | | CW1 on W | | CW2 on W | | CW1 on X | | CW2 on X | | CW1 on W | | CW2 on W | |
| Number of Indicators | Quality of Indicators | Number of Sites | Site Size | Bias | Type I error | Bias | Type I error | Bias | Type I error | Bias | Type I error | Bias | Type I error | Bias | Type I error | Bias | Power | Bias | Power | Bias | Power | Bias | Power | Bias | Type I error | Bias | Type I error | Bias | Power | Bias | Power | Bias | Power | Bias | Power |
| 6 | 0.7 | 50 | 30 | 0.006 | 0.048 | -0.004 | 0.123 | -0.010 | 0.064 | -0.007 | 0.131 | -0.005 | 0.032 | -0.010 | 0.094 | 0.031 | 0.569 | 0.075 | 0.992 | 0.012 | 0.577 | 0.054 | 1.000 | 0.001 | 0.042 | -0.006 | 0.101 | 0.020 | 0.731 | 0.065 | 0.998 | 0.019 | 0.651 | 0.053 | 0.998 |
| 6 | 0.7 | 50 | 60 | 0.005 | 0.046 | 0.001 | 0.090 | 0.003 | 0.064 | -0.003 | 0.080 | -0.005 | 0.048 | 0.003 | 0.132 | 0.006 | 0.840 | 0.021 | 0.998 | 0.005 | 0.894 | 0.027 | 1.000 | -0.003 | 0.052 | 0.000 | 0.112 | 0.009 | 0.964 | 0.027 | 1.000 | 0.012 | 0.896 | 0.022 | 1.000 |
| 6 | 0.7 | 150 | 30 | -0.004 | 0.048 | -0.003 | 0.106 | 0.001 | 0.032 | 0.003 | 0.072 | 0.000 | 0.044 | -0.004 | 0.100 | 0.005 | 0.974 | 0.022 | 1.000 | 0.006 | 0.982 | 0.017 | 1.000 | 0.000 | 0.022 | -0.002 | 0.094 | 0.009 | 0.998 | 0.015 | 1.000 | 0.006 | 0.990 | 0.013 | 1.000 |
| 6 | 0.7 | 150 | 60 | 0.000 | 0.038 | 0.005 | 0.104 | -0.001 | 0.036 | 0.001 | 0.074 | 0.003 | 0.054 | 0.002 | 0.120 | 0.004 | 1.000 | 0.011 | 1.000 | 0.000 | 1.000 | 0.000 | 1.000 | 0.001 | 0.030 | 0.002 | 0.062 | 0.004 | 1.000 | 0.012 | 1.000 | 0.000 | 0.996 | 0.010 | 1.000 |
| 6 | 0.8 | 50 | 30 | 0.005 | 0.052 | 0.000 | 0.114 | -0.008 | 0.066 | 0.001 | 0.082 | -0.006 | 0.064 | -0.006 | 0.106 | 0.011 | 0.942 | 0.021 | 1.000 | 0.007 | 0.980 | 0.016 | 1.000 | 0.001 | 0.050 | -0.007 | 0.084 | 0.009 | 0.982 | 0.017 | 1.000 | 0.005 | 0.916 | 0.017 | 1.000 |
| 6 | 0.8 | 50 | 60 | 0.002 | 0.070 | -0.001 | 0.110 | 0.000 | 0.094 | -0.002 | 0.072 | -0.002 | 0.072 | 0.001 | 0.130 | -0.001 | 0.988 | 0.002 | 0.998 | 0.002 | 1.000 | 0.007 | 1.000 | -0.002 | 0.066 | 0.001 | 0.070 | 0.001 | 1.000 | 0.004 | 1.000 | 0.006 | 0.990 | 0.003 | 1.000 |
| 6 | 0.8 | 150 | 30 | -0.003 | 0.048 | -0.002 | 0.104 | 0.001 | 0.044 | 0.003 | 0.052 | -0.001 | 0.058 | -0.002 | 0.074 | 0.000 | 1.000 | 0.010 | 1.000 | -0.001 | 1.000 | 0.003 | 1.000 | 0.000 | 0.024 | -0.003 | 0.082 | 0.002 | 1.000 | 0.003 | 1.000 | 0.002 | 1.000 | 0.005 | 1.000 |
| 6 | 0.8 | 150 | 60 | 0.000 | 0.056 | 0.003 | 0.108 | 0.001 | 0.062 | 0.003 | 0.064 | 0.001 | 0.072 | 0.001 | 0.072 | 0.001 | 1.000 | 0.004 | 1.000 | -0.003 | 1.000 | -0.003 | 1.000 | -0.001 | 0.018 | 0.000 | 0.070 | 0.001 | 1.000 | 0.004 | 1.000 | -0.002 | 1.000 | 0.003 | 1.000 |
| 6 | 0.9 | 50 | 30 | 0.006 | 0.068 | 0.002 | 0.068 | -0.005 | 0.074 | 0.003 | 0.088 | -0.003 | 0.076 | -0.003 | 0.092 | 0.008 | 0.984 | 0.015 | 1.000 | 0.004 | 0.998 | 0.009 | 1.000 | 0.003 | 0.066 | -0.006 | 0.066 | 0.004 | 1.000 | 0.009 | 1.000 | 0.003 | 0.954 | 0.014 | 1.000 |
| 6 | 0.9 | 50 | 60 | 0.002 | 0.076 | -0.002 | 0.078 | -0.001 | 0.088 | 0.000 | 0.088 | -0.001 | 0.078 | 0.003 | 0.104 | -0.001 | 0.994 | 0.004 | 1.000 | 0.000 | 1.000 | 0.002 | 1.000 | -0.003 | 0.080 | 0.001 | 0.044 | 0.002 | 1.000 | 0.003 | 1.000 | 0.004 | 0.992 | 0.005 | 0.998 |
| 6 | 0.9 | 150 | 30 | -0.003 | 0.066 | -0.002 | 0.074 | 0.001 | 0.040 | 0.001 | 0.054 | -0.001 | 0.054 | -0.004 | 0.074 | 0.001 | 1.000 | 0.004 | 1.000 | -0.001 | 1.000 | 0.002 | 1.000 | 0.001 | 0.028 | -0.001 | 0.074 | 0.000 | 1.000 | 0.001 | 1.000 | 0.002 | 1.000 | 0.003 | 1.000 |
| 6 | 0.9 | 150 | 60 | 0.000 | 0.076 | 0.002 | 0.090 | 0.001 | 0.054 | 0.002 | 0.044 | 0.001 | 0.072 | 0.001 | 0.084 | 0.000 | 1.000 | 0.001 | 1.000 | -0.002 | 1.000 | -0.003 | 1.000 | 0.000 | 0.038 | -0.002 | 0.058 | 0.000 | 1.000 | 0.001 | 1.000 | -0.003 | 1.000 | 0.001 | 1.000 |
| 12 | 0.7 | 50 | 30 | 0.002 | 0.050 | -0.002 | 0.086 | 0.009 | 0.078 | 0.007 | 0.084 | -0.002 | 0.048 | 0.000 | 0.116 | 0.012 | 0.890 | 0.015 | 0.998 | 0.016 | 0.960 | 0.026 | 1.000 | -0.001 | 0.060 | 0.003 | 0.098 | 0.004 | 0.944 | 0.017 | 1.000 | 0.004 | 0.886 | 0.011 | 1.000 |
| 12 | 0.7 | 50 | 60 | -0.002 | 0.068 | -0.002 | 0.104 | -0.010 | 0.092 | -0.001 | 0.102 | 0.006 | 0.084 | 0.013 | 0.102 | 0.000 | 0.978 | 0.010 | 1.000 | 0.003 | 1.000 | 0.008 | 1.000 | -0.004 | 0.074 | 0.000 | 0.076 | 0.003 | 1.000 | 0.014 | 1.000 | 0.005 | 0.980 | 0.020 | 1.000 |
| 12 | 0.7 | 150 | 30 | -0.003 | 0.048 | -0.002 | 0.064 | 0.003 | 0.048 | -0.001 | 0.044 | 0.002 | 0.032 | 0.003 | 0.092 | 0.002 | 1.000 | 0.004 | 1.000 | 0.001 | 1.000 | 0.006 | 1.000 | 0.003 | 0.040 | 0.002 | 0.076 | 0.003 | 1.000 | 0.010 | 1.000 | 0.006 | 0.998 | 0.009 | 1.000 |
| 12 | 0.7 | 150 | 60 | 0.001 | 0.052 | 0.003 | 0.088 | -0.003 | 0.046 | -0.004 | 0.056 | -0.001 | 0.048 | 0.000 | 0.076 | -0.001 | 1.000 | 0.002 | 1.000 | -0.003 | 1.000 | 0.002 | 1.000 | 0.002 | 0.044 | -0.001 | 0.048 | 0.002 | 1.000 | 0.003 | 1.000 | 0.000 | 1.000 | 0.003 | 1.000 |
| 12 | 0.8 | 50 | 30 | 0.004 | 0.066 | 0.003 | 0.084 | 0.007 | 0.092 | 0.001 | 0.068 | -0.001 | 0.052 | 0.002 | 0.118 | 0.008 | 0.974 | 0.011 | 1.000 | 0.010 | 0.998 | 0.017 | 1.000 | -0.002 | 0.072 | 0.003 | 0.062 | 0.002 | 0.998 | 0.013 | 1.000 | -0.001 | 0.967 | 0.004 | 1.000 |
| 12 | 0.8 | 50 | 60 | -0.003 | 0.082 | -0.003 | 0.106 | -0.005 | 0.094 | 0.001 | 0.098 | 0.004 | 0.068 | 0.006 | 0.092 | 0.001 | 0.994 | 0.004 | 1.000 | 0.003 | 1.000 | 0.004 | 1.000 | -0.002 | 0.056 | -0.002 | 0.072 | 0.002 | 1.000 | 0.007 | 1.000 | 0.004 | 0.994 | 0.012 | 1.000 |
| 12 | 0.8 | 150 | 30 | -0.003 | 0.062 | -0.003 | 0.104 | 0.002 | 0.048 | -0.002 | 0.034 | 0.001 | 0.038 | 0.001 | 0.070 | -0.001 | 1.000 | 0.000 | 1.000 | 0.002 | 1.000 | 0.006 | 1.000 | 0.002 | 0.044 | 0.000 | 0.058 | 0.001 | 1.000 | 0.004 | 1.000 | 0.003 | 1.000 | 0.004 | 1.000 |
| 12 | 0.8 | 150 | 60 | 0.001 | 0.058 | 0.003 | 0.096 | -0.002 | 0.062 | -0.001 | 0.058 | -0.001 | 0.042 | -0.001 | 0.088 | -0.002 | 1.000 | 0.000 | 1.000 | -0.002 | 1.000 | 0.000 | 1.000 | 0.001 | 0.050 | 0.001 | 0.050 | 0.001 | 1.000 | 0.001 | 1.000 | -0.001 | 1.000 | -0.001 | 1.000 |
| 12 | 0.9 | 50 | 30 | 0.003 | 0.056 | 0.002 | 0.096 | 0.006 | 0.098 | -0.001 | 0.066 | -0.001 | 0.050 | 0.001 | 0.100 | 0.007 | 0.986 | 0.011 | 0.998 | 0.007 | 1.000 | 0.013 | 1.000 | -0.003 | 0.080 | 0.001 | 0.064 | 0.001 | 1.000 | 0.008 | 1.000 | 0.000 | 0.984 | 0.002 | 1.000 |
| 12 | 0.9 | 50 | 60 | -0.003 | 0.066 | -0.002 | 0.092 | -0.005 | 0.092 | 0.001 | 0.112 | 0.005 | 0.088 | 0.006 | 0.094 | 0.002 | 1.000 | 0.004 | 1.000 | 0.003 | 1.000 | 0.005 | 1.000 | -0.004 | 0.054 | -0.004 | 0.066 | 0.002 | 1.000 | 0.006 | 1.000 | 0.002 | 0.996 | 0.009 | 1.000 |
| 12 | 0.9 | 150 | 30 | -0.003 | 0.058 | -0.003 | 0.080 | 0.001 | 0.060 | -0.001 | 0.044 | 0.001 | 0.042 | 0.001 | 0.078 | -0.001 | 1.000 | 0.000 | 1.000 | 0.001 | 1.000 | 0.005 | 1.000 | 0.002 | 0.044 | 0.000 | 0.058 | -0.001 | 1.000 | 0.002 | 1.000 | 0.001 | 1.000 | 0.003 | 1.000 |
| 12 | 0.9 | 150 | 60 | 0.000 | 0.066 | 0.003 | 0.078 | -0.002 | 0.068 | 0.000 | 0.062 | 0.000 | 0.052 | 0.000 | 0.102 | -0.001 | 1.000 | 0.000 | 1.000 | -0.002 | 1.000 | -0.001 | 1.000 | 0.002 | 0.054 | 0.001 | 0.056 | 0.001 | 1.000 | 0.001 | 1.000 | -0.001 | 1.000 | -0.002 | 1.000 |

*4.3.2 Classification Accuracy*

The classification error rate of NP-MLCA was evaluated by calculating the proportion of individuals whose predicted class membership was not the same as their true class membership. Similarly, the classification error rate at the site level was evaluated by calculating the proportion of sites whose predicted class membership was not the same as their true class membership. Figures 4 and 5 present individual-level and site-level classification error rates using six interaction plots.

The trends of classification rates in Figure 4 and Figure 5 are very similar, but the differences within study factors showed a more noticeable difference in terms of level-2 classification error. For example, comparing Figure 4a to 5a and Figure 4d to 5d, it is apparent that increasing the number of sites lowered the level-2 classification error rate, but not the level-1 classification error rate. Comparing Figure 4b to 5b, Figure 4e to 5e, and Figure 4f to 5f, it can be seen that increasing the site size brought down the level-2 classification error rate.

Both level-1 and level-2 classification error rates were affected by CRP and the number of indicators. However, Figure 5 showed that, in addition to CRP and the number of indicators, the number of sites and the site size should also be considered as the influential factors to the level-2 classification error rates. Thus, as long as the quality of indicators are [0.2, 0.8] or more distinct (i.e., [0.1, 0.9]), even under the small size scenario of 50 sites and site size 30, the classification accuracy rate at the level-1 level of multilevel latent class model can exceed 80%. If researchers are more interested in level-2 classification accuracy, many study factors, such as site size, number of indicators, number of sites, and the quality of indicators, should be taken into consideration when designing the research.

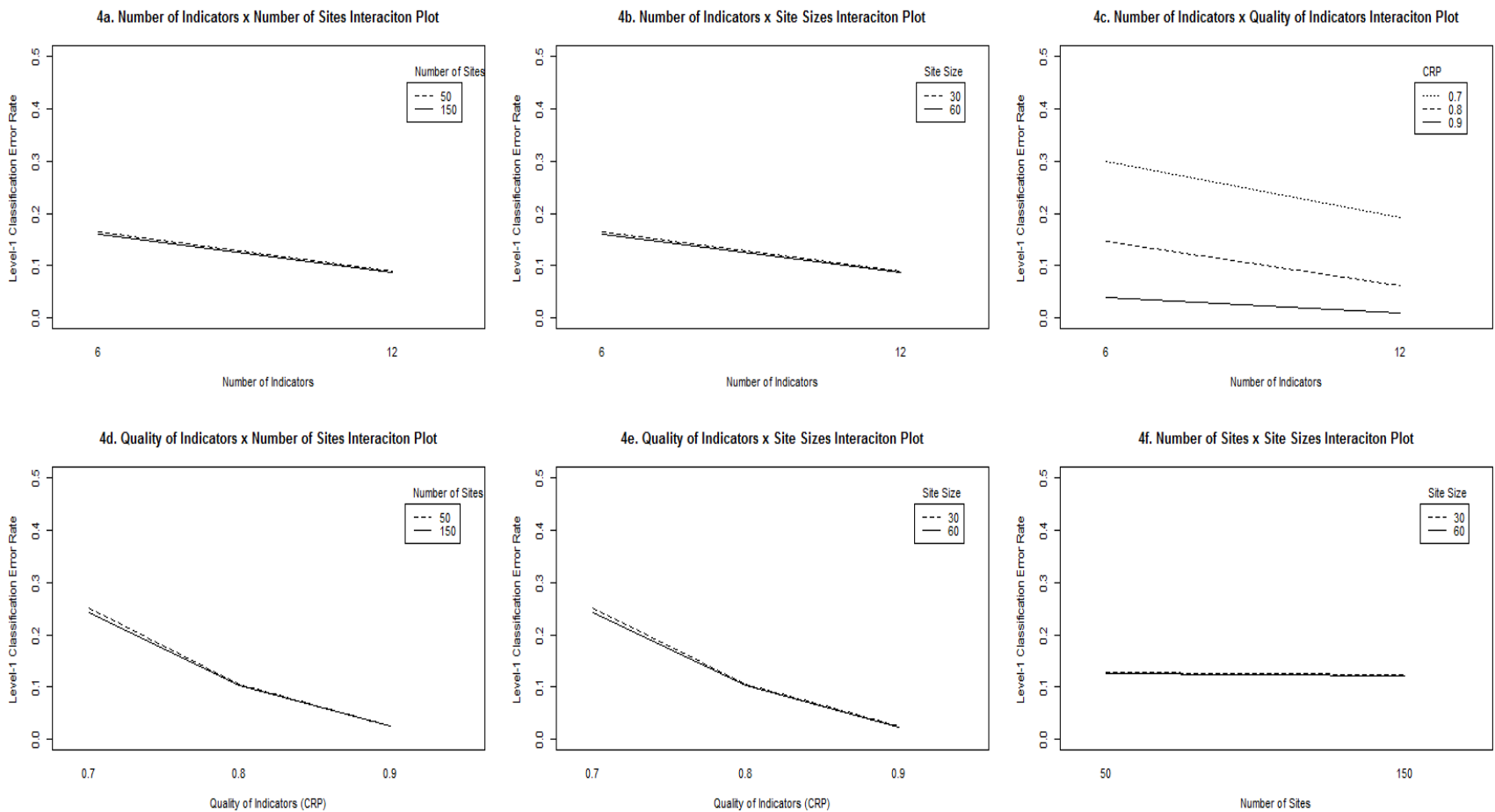

*Figure 4.* The Interaction Plots of Level-1 Classification Error Rate

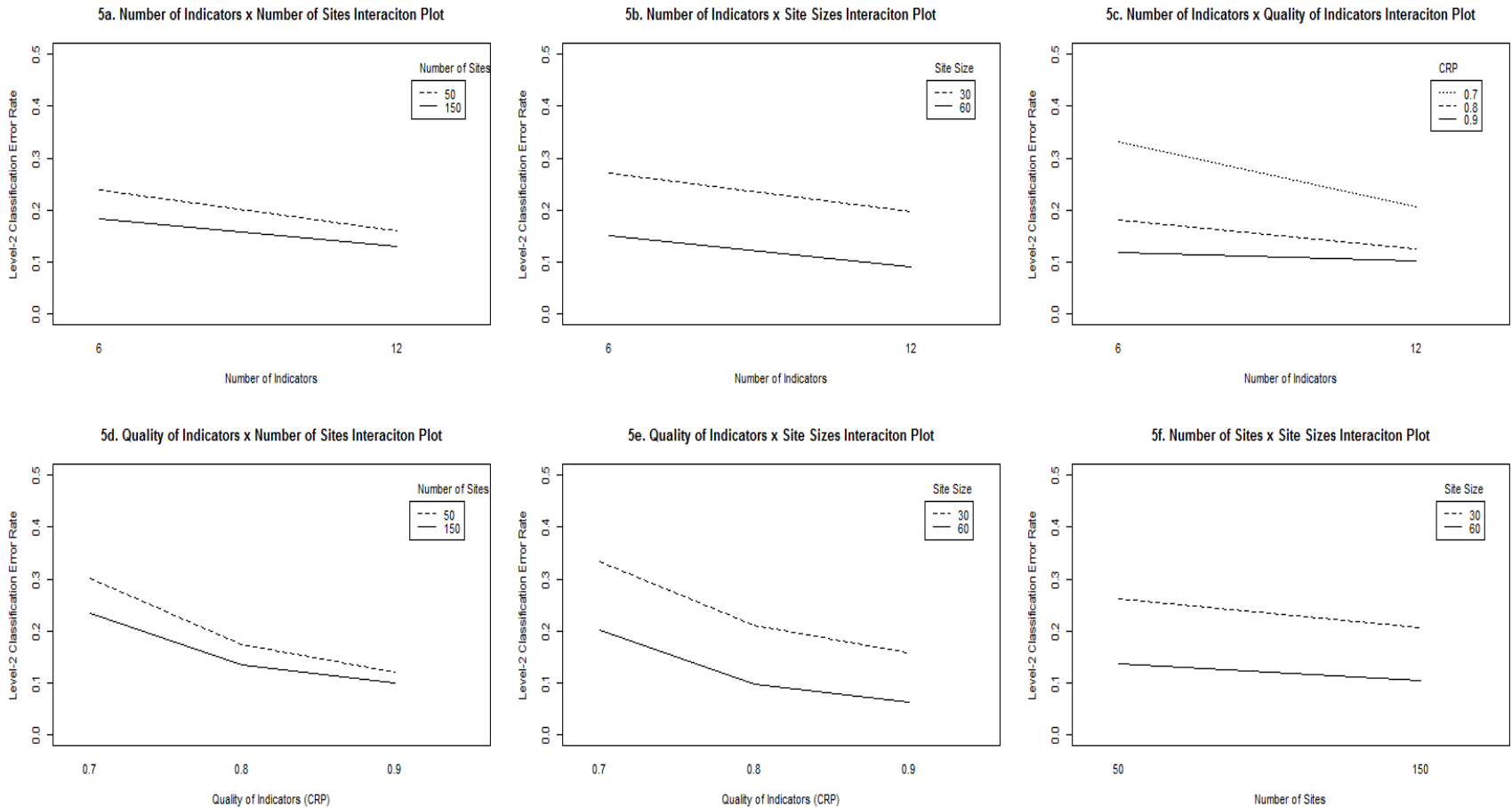

*Figure 5.* The Interaction Plots of Level-2 Classification Error Rate

We further conducted two ANOVA models to examine the effect of the six study factors (i.e., number of sites, site size, number of indicators, quality of indicators, level-1 covariate effect, and cross-level covariate effect) on the level-1 classification error rate and level-2 classification error rate. The results showed that none of the two-way interactions of 6 study factors had an effect on the classification rate. The quality of the indicators had a strong effect ($\eta^2 = 0.815$), the number of indicators had a medium effect ($\eta^2 = 0.130$), and the site size had a small effect ($\eta^2 = 0.029$) on decreasing the level-1 classification error. The cross-level effect had a strong effect ($\eta^2 = 0.855$), and both site size and quality of indicators had small effects ($\eta^2 = 0.043, 0.024$, respectively) on decreasing the site-level (i.e., level-2) classification error rate.

Overall, the parameter recovery of the covariate estimates and CRP estimates showed great performance in parameter estimates and standard error estimates. The 'bottom limit' condition that we designed in the simulation scenarios: models with 6 indicators, [0.3, 0.7] CRP, 50 sites, and 30 individuals in each site, does show little biases estimates and standard errors. As for the classification accuracy at level-1 and level-2, the result showed that increasing the quality of indicators increased the level-1 classification accuracy, while multiple factors affected level-2 classification accuracy, which included the cross-level effects, site size, and quality of indicators.

## 5. Discussion

Classification accuracy and covariate effects in the multilevel context are the main interests of this study. In addition to the varying levels of the study factors designed in the simulation conditions, we included both consistent-response patterns and mixed-response patterns in the latent classes to make the simulation scenarios closer to the real world. A total of 96 conditions was investigated to answer the research question: how are classification accuracy and parameter estimation of nonparametric multilevel latent class modeling affected by six study factors: 1) the conditional response probabilities (i.e., the quality of indicators), 2) the number of latent class indicators, 3) the level-1 covariate effects, 4) the cross-level covariate effects, 5) the number of sites (i.e., level-2 units), and 6) the site size (i.e., the number of level-1 units in each level-2 unit).

The findings of the simulation regarding level-1 classification accuracy were consistent with previous studies (Finch & French, 2013; Wurpts & Geiser, 2014; Yu & Park, 2014). Increasing the number of indicators and CRP improved the classification accuracy rate. When the CRP is 0.8 or higher (or 0.2 and lower), the classification accuracy rate reached 80% and above. Level-1 classification accuracy was not affected by the number of indicators, number of sites, site size, or the cross-level covariate effect. It was only when the separation of the classes was not distinct (i.e., CRP [0.3, 0.7]) and only 6 indicators in the model that the student-level classification accuracy would be problematic (lower than 80%). In short, increasing the number of sites or the size of sites did not affect the level-1 classification accuracy, but increasing the number of indicators and the level-1 latent class separation improved the level-1 latent class classification accuracy. This is not far from our expectation, since the classification depends on how distinctly classes were separated and how much information we gathered. This also explains CRP's alternative names: the quality of indicators or latent class distinctiveness. After all, it is the CRPs that defined the latent classes in the measurement model. The more separated the classes were, the more easily the class membership could be recovered, and the better the classification accuracy would be. Similarly, the greater the number of indicators involved in the model, the easier it was for the level-1 class membership to be identified correctly. Note that the smallest sample size in the current study is 1,500.

While the number of sites and size of sites had no impact on the level-1 classification accuracy, increasing either the number of sites or the size of the sites improved the level-2 classification accuracy. This finding is consistent with Finch & French (2013), though the simulation data in their study was generated from parametric multilevel latent class analysis and modeled by nonparametric multilevel latent class analysis. The results of the current study show that the differences of level-2 classification accuracy among the conditions (CRP [0.3, 0.7], [0.2, 0.8] and [0.1, 0.9]) were not as large as those of the level-1 classification accuracy, but the number of sites and the size of sites had an impact on the level-2 classification accuracy. Again, CRP [0.3, 0.7] and 6 indicators exhibited a problem with the level-2 classification accuracy, with accuracy as low as 70%. This result is expected from the structure of NP-MLCA: the level-1 classes are nested within level-2 classes. Since the contribution to the level-2 classification was solely from level-1 classes, level-2 classification accuracy can be sensitive not only to the factors that affect the level-1 classification accuracy, but to the size-related factors. Thus, the quality of the indicators, the cross-level covariate effects, and the sample sizes could affect level-2 classification accuracy via level-1 latent classes. Therefore, it is no surprise that if the level-1 classification accuracy did not perform well in the condition CRP [0.3, 0.7], the level-2 classification accuracy was harmed by the low quality of the indicators, as well. In addition, applied researchers can learn from Figures 5 and 6 that if one cannot find indicators with more separated CRP than [0.3, 0.7], increasing the number of indicators to 12 can improve the classification accuracy.

In comparison to other conditions in the current study, the parameter recovery of CRP estimates across all three latent classes were slightly problematic when the condition included only 6 indicators, CRP [0.3, 0.7], 50 sites, site size 30, and nuisance variables at both levels.

However, the results were comparable with Park & Yu (2018a). Although the biases of the CRP estimates under these conditions were not serious, our findings suggest that it is crucial to choose covariates that have theoretical support, and to have a covariate selection process before running the conditional NP-MLCA model. Especially when the quality of indicators is low and the sample size is small, including nuisance covariates can be detrimental to the CRP estimates. Either increasing the sample size, the quality of indicators, and/or the number of indicators can improve the parameter recovery performance of the CRP estimates.

The parameter recovery of cross-level covariate effects has not been studied previously. Previous simulation studies aggregated results from $k$-1 covariate estimates to evaluate the parameter recovery if there were $k$ latent classes in the model, while $k$-1 covariate effects were supposed to be reported for each covariate. Ignoring important differences between the estimates may lead to inaccurate results. Our findings showed that there are differences in the results of covariate parameter recovery on different latent classes when the covariates have nuisance effects.

The power to detect first-level and cross-level covariate effects in nonparametric MLCAs has not been studied previously. The finding showed that the power reached 0.992 or higher to detect a strong first-level effect and cross-level effect in nonparametric MLCA, even in the conditions with the smallest sample size (i.e., 1,500), the quality of indicators [0.3, 0.7], and a mixed-response pattern latent class as the outcome. As for detecting a small covariate effect in the model, the minimum required sample size for 80% power is 3,000.

When nuisance covariates were included in the model, the Type-I error rates of the covariate estimates were found to be inflated. This finding is not surprising since it is known that the covariates' standard error estimates under the ML method tend to be underestimated overall in a single-level latent class model (Z. Bakk, Oberski, & Vermunt, 2014; Zsuzsa Bakk et al., 2013; Vermunt, 2010). With unbiased estimates, it is expected that the underestimated standard errors would lead to inflated Type-I errors of the estimates. However, one thing worth noting in this study is that we found the inflations only occurred when the mixed response pattern latent classes were regressed. When the consistent-response pattern latent class was regressed on nuisance covariates, the Type-I error of the effect was around 0.05. The changes of Type-I errors did not seem to be attributed to the other study factors, such as site size, number of sites, CRP, number of indicators, or the level of the covariate. Since the reference latent class in this study is also in the consistent-response pattern, whether the systematic bias in standard error estimate is specific to the type of response pattern in the latent class may be worth investigating in future studies.

## 6. Conclusion

From the findings of this study, we can conclude that as long as the covariates are not nuisance variables, incorporating strong-effect covariates at the first-level and/or at the second-level can improve CRP estimation and higher-level classification accuracy, even when the model has low quality indicators and a small number of indicators. The number of indicators and the class distinctiveness (i.e. CRP) affects the level-1 latent class accuracy. For higher-level latent class classification accuracy, incorporating a second-level covariate with strong cross-level effect and increasing the site size and number of sites can be beneficial. This finding can benefit applied classification researchers when good quality indicators are difficult to find.

The label switching issue is one of the commonly cited limitations of mixture model simulation studies. Although we followed previous research and took care of all switched cases, identifying the switching combination to which each output belongs was somewhat difficult, especially when the latent class was less distinctive (i.e., true CRP parameter was [0.3, 0.7] ) and mixed response pattern latent classes were involved. In the current study, among 500 replications, the proportion of switched cases ranged from 20% to 60% across conditions. A detailed explanation of the potential complexities of the simulation procedure for latent class analysis can be found in Chang (2019).

We varied the study factors to investigate model performance under various conditions. Expanding from these conditions, future studies can investigate the performance of NP-MLCA under model misspecification. In addition, a three-step approach to estimation for the cross-level covariate effect is worth investigating in the future to provide an alternative option to contextualize the latent classes.

The present study is an important step toward understanding the classification accuracy and parameter recovery of CRP, cross-level covariate effects, and level-1 covariate effects under various conditions. The significant difference between this study and previous studies is that the data was generated from nonparametric multilevel models instead of parametric multilevel models. As applied research using NP-MLCA is booming, the findings of this study should provide some guidelines and help foster discussion for applied researchers who are interested in simultaneous classification at different levels in multilevel contexts.